\documentclass[11pt]{article}
\usepackage{graphicx}
\usepackage{amsmath}
\input psfig.sty

\newlength{\back}
\newlength{\greater}

\begin{document}

\begin{titlepage}

\title{Spontaneous Symmetry Breaking, Non-minimal Coupling,\\
and Cosmological Constant Problem\\
\vspace{0.5cm}}

\author{Je-An \ Gu\thanks{%
E-mail address: wyhwange@phys.ntu.edu.tw} \ \ \ and \ \ W-Y. P. Hwang\thanks{%
E-mail address: wyhwang@phys.ntu.edu.tw} \\
{\small Department of Physics, National Taiwan University, Taipei
106, Taiwan, R.O.C.}
\medskip
}

\date{\small \today}

\maketitle

\begin{abstract}
Treating the gravitational field as a dynamical field, we study
the spontaneous symmetry breaking induced by a scalar field under
its self-interaction and non-minimal interaction with gravity in
four dimensional space-time. In particular, we explore the
feasibility of inducing spontaneous symmetry breaking after
introducing the non-minimal coupling, and discuss briefly the
cosmological constant problem corresponding to the phase
transition associated with the spontaneous symmetry breaking.
\end{abstract}

\vspace{9.7cm}

\begin{flushleft}
\end{flushleft}

\end{titlepage}

\section{Introduction}
In physics, the standard way of implementing spontaneous symmetry
breaking (SSB) is by introducing a scalar field with a potential
like
\begin{equation}
V(\Phi )=-\frac{1}{2}\mu ^{2}\Phi ^{2}+\frac{1}{4}\lambda \Phi
^{4}, %
\label{Higgs potential}
\end{equation}
which entails multiple true vacua (at $|\Phi | = \sqrt{\mu ^2 /
\lambda }$).\footnote{A famous example is the Higgs mechanism in
the Standard Model to break the SU(2)$\times $U(1) gauge symmetry
spontaneously \cite{Higgs etc}.}  In the process of choosing one
among these true vacua, SSB may be induced. We note that the
negative $(mass)^{2}$ $\Phi ^{2}$ term and $\lambda \Phi ^{4}$
term in Eq.\ (\ref{Higgs potential}) together play a crucial role
in making the potential entail multiple true vacua. As the scalar
field rolls down to one of the true vacua from the origin, some
phase transition may happen, and accordingly, its ``latent heat''
(with the amount of $|\delta V |= \mu ^2 / (4 \lambda )$) will be
released to the ``thermal energy''. In the framework of quantum
field theory, the ``latent heat'' is the difference of vacuum
energy in different phases, and the ``thermal energy'' is the
energy of particles produced along with the phase transition. As
pointed out by Zeldovich \cite{Zeldovich:1968} (see also
\cite{Weinberg:1989cp}), the vacuum energy will contribute to the
effective cosmological constant. Thus, if we need a comparatively
small effective cosmological constant (for the indication from
various astronomical observations, for instance, the supernova
distance measurements \cite{Perlmutter:1999np,Riess:1998cb}) after
the phase transition, we must fine tune the parameters: $\mu $ and
$\lambda $, and the original cosmological constant $\Lambda_0 $
(before the phase transition), such that
\begin{equation}
\frac{1}{\kappa }\Lambda_0 \simeq \frac{\mu^4}{4 \lambda} , %
\label{fine tunning condition 1}
\end{equation}
where $\kappa \equiv 8\pi G$, and $G$ is the gravitational
constant. Such a fine tuning is one of various versions of the
cosmological constant problem.

The \emph{non-minimal coupling} between scalar fields and gravity
has been introduced in various topics in cosmology (see
\cite{Sahni:1998at,Flachi:2000ca} and references therein) and
quantum field theories (for a review, see \cite{Buchbinder:1992rb}
and references therein). In the framework of quantum field theory
in curved space-time, the non-minimal coupling can be introduced
by the requirement of renormalizability \cite{Buchbinder:1984xx}
(for a review, see \cite{Buchbinder:1989zz}). In particular, the
role of the non-minimal coupling in SSB and phase transitions in
four dimensional space-time has been widely discussed (for a
review, see \cite{Buchbinder:1992rb}): That the non-minimal
coupling with the `external' gravitational field may lead to SSB
has been pointed out \cite{Janson1976&Grib1977}. It has also been
noted that SSB and phase transitions can be induced by curvature
via the non-minimal coupling with the `external' gravitational
field \cite{R-induced SSB}, and moreover, the curvature can be a
symmetry-breaking factor \cite{SB&VacuumStabi}. We note that, in
the work mentioned above, the effect of the curved space-time is
taken into account by introducing an `external' gravitational
field, that is, the metric tensor of the space-time is treated as
a background. This is different from the usage of the
gravitational field in our recent work to be discussed below.

Recently we study the influence of the non-minimal coupling with
the \emph{dynamical} gravitational field on SSB in the space-time
of an arbitrary dimension \cite{Gu:2001rr}, where we keep only the
negative $(mass)^{2}$ $\Phi ^{2}$ term, but discard $\lambda \Phi
^{4}$ term in the potential (Eq.~(\ref{Higgs potential}))used in
the standard SSB scenario. In this paper, we take \emph{both} two
terms in Eq.~(\ref{Higgs potential}) into consideration, and focus
on \emph{four} dimensional space-time. In particular, we will
consider a universe which is dominated by a scalar field and
possibly by a cosmological constant, and show the feasibility of
inducing SSB in this setup. We will also discuss briefly the
cosmological constant problem corresponding to the phase
transition associated with SSB.

\section{The Basics}
We consider a four dimensional universe, which is dominated by a
scalar field, a cosmological constant, and gravity, and described
by the following action:
\begin{equation}
S = -\frac{1}{2\kappa }\int d^{4}x\sqrt{g}\left(
    \mathcal{R}+2\Lambda \right)
    + \int d^{4}x\sqrt{g}\mathcal{L}_{\Phi }, %
\label{action}
\end{equation}
where $\kappa \equiv 8\pi G$ ($G$: gravitational constant), $g$ is
the absolute value of the determinant of the metric tensor
$g_{\alpha \beta}$, $\Lambda $ is the cosmological constant, and
$\mathcal{L}_{\Phi }$ is the Lagrangian density of the scalar
field $\Phi $. We would like to introduce the non-minimal coupling
into the scalar-field potential (or Lagrangian density
$\mathcal{L}_{\Phi }$), including both two terms in Eq.\
(\ref{Higgs potential}) introduced in the standard SSB scenario.
Accordingly, we consider
\begin{equation}
\mathcal{L}_{\Phi }=\frac{1}{2}g^{\alpha \beta}\left( \partial
_{\alpha}\Phi \right) \left( \partial _{\beta}\Phi \right)
-V_{\mathcal{R}}\left( \Phi \right) , \quad \alpha ,\beta =0,1,2,3
\end{equation}
\begin{equation}
V_{\mathcal{R}}\left( \Phi \right) = \frac{1}{2}\xi \mathcal{R}
\Phi ^{2}-\frac{1}{2}\mu ^{2}\Phi ^{2}+\frac{1}{4}\lambda \Phi^4 ,
\label{potential including non-minimal coupling}
\end{equation}
where the term $\frac{1}{2}\xi \mathcal{R} \Phi ^2$ in the
scalar-field potential $V_{\mathcal{R}}(\Phi )$ is the non-minimal
coupling term, which couple the scalar field $\Phi $ with gravity
via the Ricci scalar $\mathcal{R}$ with a coupling constant $\xi $
to be positive in our consideration.\footnote{The convention for
the metric tensor in this paper is:
\[
g _{\alpha \beta}=\left( +,-,-,- \right).
\]
}

The variation of the action in Eq.\ (\ref{action}) with respect to
the scalar field $\Phi $ and the metric tensor $g^{\alpha \beta}$
yields the field equation of $\Phi $:
\begin{equation}
\Phi^{;\gamma }_{\;\;\, ;\gamma }+\left( \xi \mathcal{R}-\mu ^{2}
+ \lambda \Phi ^2 \right) \Phi =0, \label{field equation of Phi}
\end{equation}
and the Einstein equations:
\begin{equation}
G_{\alpha \beta} - \Lambda g_{\alpha \beta} = \kappa \left\{ %
\left( \partial _{\alpha}\Phi \right) \left( \partial _{\beta}\Phi
\right) - \mathcal{L}^{\left( 0 \right) }_{\Phi }g_{\alpha \beta}
-\xi \Phi ^{2} G_{\alpha \beta} + \xi \left( \Phi ^{2} \right)
_{;\alpha ;\beta} - \xi \left( \Phi ^{2}
\right)^{;\gamma}_{\;\;\, ;\gamma} g_{\alpha \beta} \right\} , %
\label{Einstein equation}
\end{equation}
where
\begin{equation}
\mathcal{L}^{\left( 0 \right) }_{\Phi } = %
\frac{1}{2} g^{\alpha ' \beta '} %
\left( \partial _{\alpha '} \Phi \right) %
\left( \partial _{\beta '} \Phi \right) + %
\frac{1}{2} \mu ^{2} \Phi ^{2} - \frac{1}{4}\lambda \Phi^4 ,
\end{equation}
$G_{\alpha \beta}$ is the Einstein tensor, and the semicolon `;'
denotes the `covariant derivative'. (We note that the non-minimal
coupling term is not included in $\mathcal{L}^{\left( 0 \right)
}_{\Phi }$.) The term $-\xi \Phi ^{2}G_{\alpha \beta}$ in the
brace in Eq.\ (\ref{Einstein equation}) will modify the Einstein
equations, the gravitational constant $G$ (also $\kappa $), and
the cosmological constant $\Lambda $ as follows:
\begin{equation}
G_{\alpha \beta} - \Lambda ' g_{\alpha \beta} = \kappa _{eff} \left\{ %
\left( \partial _{\alpha}\Phi \right) \left( \partial _{\beta}\Phi
\right) - \mathcal{L}^{\left( 0 \right) }_{\Phi }g_{\alpha \beta}
+ \xi \left( \Phi ^{2} \right) _{;\alpha ;\beta} %
- \xi \left( \Phi ^{2} \right)^{;\gamma}_{\;\;\, ;\gamma}
g_{\alpha \beta} \right\}, \label{modified Einstein equation}
\end{equation}
where the `modified cosmological constant' $\Lambda '$ and the
`effective gravitational constant' $G_{eff}$, as well as $\kappa
_{eff}$, are defined by
\begin{eqnarray}
\Lambda '&\equiv &\frac{\Lambda }{1+\kappa \xi \Phi ^{2}} \\
8\pi G_{eff} &\equiv& \kappa _{eff} \equiv \frac{\kappa }{1+\kappa
\xi \Phi ^{2}}.
\end{eqnarray}

Taking trace of both sides of the \emph{modified} Einstein
equations (\ref{modified Einstein equation}) gives us the Ricci
scalar $\mathcal{R}$ as a function of the scalar field $\Phi $ and
its covariant derivatives:
\begin{equation}
\mathcal{R} = \frac{\kappa }{1 + \kappa \xi \Phi ^{2}} \left\{ %
-4 \left( \frac{1}{\kappa }\Lambda - \frac{1}{2} \mu ^{2} \Phi
^{2} + \frac{1}{4}\lambda \Phi ^4 \right) + \Phi ^{;\gamma}\Phi
_{;\gamma} + 3 \xi \left( \Phi ^{2} \right) ^{;\gamma}_{\;\;\,
;\gamma} \right\} . %
\label{Ricci scalar}
\end{equation}
As shown in the above equation, the Ricci scalar $\mathcal{R}$, by
which the scalar field $\Phi $ is coupled with gravity, strongly
depends on the scalar field. Such a strong dependence will make
significant influence on the inducement of SSB to be discussed in
the next section.

\section{Multiple True Vacua}
The existence of multiple true vacua is the key element of
implementing SSB. For exploring the existence of multiple true
vacua entailed by the action in Eq.\ (\ref{action}) or the
potential $V_{\mathcal{R}}\left( \Phi \right) $ in Eq.\
(\ref{potential including non-minimal coupling}), we need to find
out the constant solutions for the scalar field $\Phi $ and
explore their stability (since a vacuum state is usually
corresponding to a stable constant-field solution).

For constant $\Phi $, the field equation of $\Phi $ (\ref{field
equation of Phi}) and \emph{modified} Einstein equations
(\ref{modified Einstein equation}) become
\begin{equation}
\left( \xi \mathcal{R}-\mu ^{2} + \lambda \Phi ^2 \right) \Phi = 0, %
\label{field equation for constant Phi}
\end{equation}
\begin{eqnarray}
G_{\alpha \beta} &=&
      \Lambda ' g_{\alpha \beta} + \kappa _{eff}
      \left( -\frac{1}{2}\mu ^{2}\Phi ^{2} + \frac{1}{4}
      \lambda \Phi ^4 \right) g_{\alpha \beta} \nonumber \\
&=&   \frac{\kappa }{1+\kappa \xi \Phi ^{2}}\left( \frac{1}{\kappa } %
      \Lambda - \frac{1}{2}\mu ^{2}\Phi ^2 +
      \frac{1}{4}\lambda \Phi^4\right) g_{\alpha \beta} \nonumber \\
&=&   \kappa _{eff}V_{eff}g_{\alpha \beta}=\Lambda _{eff}g_{\alpha \beta} \, , %
\label{Lambda dominated Einstein equation}
\end{eqnarray}
where the `\emph{effective} potential energy (or vacuum energy)'
$V_{eff}$ and `\emph{effective} cosmological constant' $\Lambda
_{eff}$ are defined by
\begin{equation}
V_{eff}=\frac{1}{\kappa }\Lambda -\frac{1}{2}\mu ^{2}\Phi ^{2} +
        \frac{1}{4}\lambda \Phi ^4 ,
\end{equation}
\begin{equation}
\Lambda _{eff} = \frac{\kappa}{1+\kappa \xi \Phi ^{2}} \left(
\frac{1}{\kappa} \Lambda -\frac{1}{2} \mu ^{2}\Phi ^{2} +
\frac{1}{4}\lambda \Phi ^4 \right) = \kappa _{eff} V_{eff}.
\end{equation}

Taking trace of both sides of Eq.\ (\ref{Lambda dominated Einstein
equation}) gives us the Ricci scalar $\mathcal{R}$ as a function
of the scalar field $\Phi $:
\begin{equation}
\mathcal{R} = - \frac{4\kappa }{\left( 1+\kappa \xi \Phi ^{2} \right) } %
\left( \frac{1}{\kappa }\Lambda - \frac{1}{2}\mu ^{2}\Phi ^{2} +
\frac{1}{4} \lambda \Phi ^4 \right) . %
\label{Ricci scalar for constant Phi}
\end{equation}
Using the above equation, Eq.\ (\ref{field equation for constant
Phi}) becomes
\begin{equation}
\left[ -\frac{4\kappa \xi}{\left( 1+\kappa \xi \Phi ^{2} \right) }
\left( \frac{1}{\kappa }\Lambda - \frac{1}{2}\mu ^{2}\Phi ^{2} +
\frac{1}{4} \lambda \Phi ^4 \right) - \mu ^{2} + \lambda \Phi ^2
\right] \Phi = 0 .
\end{equation}
Then we can obtain the constant solutions for the scalar field
$\Phi $:
\begin{equation}
\Phi = \Phi _{\left( c \right) } = 0, \, \pm v
\end{equation}
where $v$ is defined by
\begin{equation}
v^{2} = \frac{4\xi \Lambda + \mu ^2}{\lambda + \kappa \xi \mu ^2}
\quad > 0 \quad \mbox{for} \quad \xi \Lambda > - \frac{1}{4} \mu ^{2}
, \quad \mbox{if} \quad \lambda > -\kappa \xi \mu ^2. %
\label{vev}
\end{equation}
The corresponding metric tensor $g_{\alpha \beta}^{(c)}$ (the
solution(s) of the modified Einstein equations (\ref{Lambda
dominated Einstein equation}) corresponding to $\Phi = \Phi
_{(c)}$) will be the metric tensor of de Sitter or anti-de Sitter
space-time, depending on the `\emph{effective} cosmological
constant' $\Lambda _{eff}$ is positive or negative.

Now we need to explore the stability of these three constant-$\Phi
$ solutions ($\Phi = \Phi _{(c)}$, $g_{\alpha \beta} = g_{\alpha
\beta }^{(c)}$). Considering small variations around these
solutions:
\begin{equation}
\left\{
\begin{array}{l}
\Phi = \Phi _{(c)} + \delta \Phi \\
g_{\alpha \beta} = g_{\alpha \beta}^{(c)} + \delta
g_{\alpha \beta} \;\; \longrightarrow \;\; %
\mathcal{R} = \mathcal{R}_{(c)} + \delta \mathcal{R}
\end{array}
\right. ,
\end{equation}
where $\mathcal{R}_{(c)}$ is the Ricci scalar for $\Phi = \Phi
_{(c)}$, the field equation of $\Phi $ (\ref{field equation of
Phi}) becomes (up to $\mathcal{O}(\delta \Phi, \delta
\mathcal{R})$):
\begin{equation}
\left( \delta \Phi \right) ^{;\gamma}_{\;\;\, ;\gamma} %
+ \left( \xi \mathcal{R}_{(c)} - \mu ^{2}
         + 3\lambda \Phi_{(c)}^2 \right)  \delta \Phi %
+ \xi \Phi _{(c)} \delta \mathcal{R} = 0, %
\label{field equation of delta Phi with R}
\end{equation}
and the equation for the Ricci scalar $\mathcal{R}$ (\ref{Ricci
scalar}) becomes (up to $\mathcal{O}(\delta \Phi, \delta
\mathcal{R})$):
\begin{eqnarray}
\delta \mathcal{R} &=& 0 \quad \mbox{for} \quad \Phi _{(c)}=0 \, , 
                       \label{delta R at zero Phi} \\
\delta \mathcal{R} &=&
       \frac{2\kappa}{1+\kappa \xi \Phi_{(c)}^{2}}
       \left( \mu ^{2} - \lambda \Phi_{c}^2 \right) \Phi _{(c)} \delta \Phi %
       + 6 \xi \Phi _{(c)} \left( \delta \Phi \right) ^{;\gamma}_{\;\;\, ;\gamma}
       \quad \mbox{for} \quad \Phi _{(c)}=\pm v. \label{delta R at vev}
\end{eqnarray}
Using above equations and Eq.\ (\ref{Ricci scalar for constant
Phi}), we can obtain the field equations of $\delta \Phi $ from
Eq.\ (\ref{field equation of delta Phi with R}): For $\Phi
_{(c)}=0$,
\begin{equation}
\left( \delta \Phi \right)^{;\gamma}_{\;\;\, ;\gamma} + \left[ -
\left( 4 \xi \Lambda + \mu ^{2} \right) \right] \delta \Phi
= 0 , %
\label{field equation of delta Phi at zero Phi without R}
\end{equation}
where \settowidth{\greater}{ >}
\begin{equation}
-\left( 4 \xi \Lambda + \mu ^{2} \right) %
\mbox{ \raisebox{0.75ex}{$<$}\hspace{-0.95\greater}\raisebox{-0.75ex}{$>$} } %
0 \quad \mbox{for} \quad \xi \Lambda %
\mbox{ \raisebox{0.75ex}{$>$}\hspace{-0.95\greater}\raisebox{-0.75ex}{$<$} } %
- \frac{1}{4} \mu ^{2}. %
\label{conditions for zero-Phi field equation}
\end{equation}
Therefore the solution $\Phi =0$ is unstable or stable for $\xi
\Lambda $ is greater or smaller than $- \mu ^2 /4$. The condition
$\xi \Lambda > - \mu ^2 /4$ is the same as the one for $v^{2}>0$,
provided $\lambda > -\kappa \xi \mu ^2$, as shown in Eq.\
(\ref{vev}). On the other hand, for $\Phi _{(c)}=\pm v$,
\begin{equation}
\left( \delta \Phi \right) ^{;\gamma}_{\;\;\, ;\gamma} %
+ 2 \left( 1 + 6 \xi ^2 v^2 \right) ^{-1} %
\left[ \frac{\kappa \xi }{1+\kappa \xi v^2} \left( \mu ^2 -
\lambda v^2 \right) + \lambda \right] v^2 \delta \Phi = 0 ,
\end{equation}
where
\begin{equation}
2 \left( 1 + 6 \xi ^2 v^2 \right) ^{-1} %
\left[ \frac{\kappa \xi }{1+\kappa \xi v^2}
\left( \mu ^2 - \lambda v^2 \right) + \lambda \right] v^2
> 0 \quad \mbox{for} \quad \lambda > -\kappa \xi \mu ^2 . %
\label{condition for vev-Phi field equation}
\end{equation}
Therefore the solutions $\Phi = \pm v$ are stable for $\lambda >
-\kappa \xi \mu ^2$. (This condition has appeared in Eq.\
(\ref{vev}), and another condition $\xi \Lambda > -\mu ^2 /4$
should also be fulfilled in order to ensure the existence of the
solutions $\Phi = \pm v$.)

Consequently, we may conclude that, provided $\lambda > -\kappa
\xi \mu ^2$, for $\xi \Lambda < \mu ^2 /4$, there is only one
constant solution, $\Phi = 0$, which is stable and corresponding
to one true vacuum, while for $\xi \Lambda > \mu ^2 /4$, there are
one unstable constant solution, $\Phi = 0$, and two stable
constant solutions, $\Phi = \pm v$, corresponding to two true
vacua.

\section{Discussion}
We have shown that the potential $V_{\mathcal{R}}(\Phi )$,
including the non-minimal coupling with gravity and the potential
terms used in the standard scenario of the spontaneous symmetry
breaking, does entail multiple true vacua, and hence be able to
induce spontaneous symmetry breaking, under suitable conditions.
In particular, as shown in Eqs.\ (\ref{vev}), (\ref{conditions for
zero-Phi field equation}), and (\ref{condition for vev-Phi field
equation}), the conditions
\begin{equation}
\left\{ %
\begin{array}{ccl}
        \kappa &>& 0 \\
           \xi &>& 0 \\
   \xi \Lambda &>& - \mu ^2 /4 \\
       \lambda &>& - \kappa \xi \mu ^2
\end{array}
\right.
\end{equation}
should be fulfilled. It is interesting to note that the
spontaneous symmetry breaking can still be achieved even for a
negative coupling constant $\lambda$, as long as it is greater
than $-\kappa \xi \mu ^2$. This is peculiar and quite different
from the standard scenario of the spontaneous symmetry breaking.

Accompanying the spontaneous symmetry breaking, under the
influence of the non-minimal coupling with gravity, both the
gravitational constant $G$ (also $\kappa $) and cosmological
constant
$\Lambda $ will undergo a ``phase transition'': %
\settowidth{\back}{>}
\begin{eqnarray}
8\pi G \equiv \kappa 
&\mbox{\raisebox{0.9ex}{\underline{\ \ \ \ SSB\ \ \ \ }}
\hspace{-2\back}$>$}& 
8\pi G_{eff} \equiv \kappa _{eff} = \left( \frac{\mu ^2 + \lambda
/ \kappa \xi}{4 \xi \Lambda + 2 \mu ^2 + \lambda / \kappa \xi}
\right) \kappa  \\
\Lambda 
&\mbox{\raisebox{0.9ex}{\underline{\ \ \ \ SSB\ \ \ \ }}
\hspace{-2\back}$>$}& 
\Lambda _{eff}= \frac{\lambda \Lambda - \kappa \mu ^4 /4 }{\lambda
+ \kappa \xi \mu ^2}  \label{Lambda phase transition}
\end{eqnarray}
when the scalar field $\Phi $ rolls down to one of the true vacua
from the origin. It is interesting to see the large $\xi \Lambda $
limit of $\kappa _{eff}$:
\[
\xi \Lambda \gg \mu ^{2},\frac{\lambda}{\kappa \xi} \;\, , \quad
\quad \kappa _{eff} \simeq \left( \frac{\mu ^2 + \lambda /(\kappa
\xi )}{4\xi \Lambda} \right) \kappa \ll \kappa .
\]
Consequently, the effective gravitational constant $G_{eff}$ (also
$\kappa _{eff}$) after the spontaneous symmetry breaking can be as
small as desired for suitable values of the parameters: $\mu $,
$\lambda$, $\xi $, and $\Lambda $, even though the original
gravitational constant $\kappa $ could be arbitrarily large. (A
similar phenomenon has been pointed out in the ``induced gravity''
model \cite{induced gravity}.)

As for the effective cosmological constant $\Lambda _{eff}$ in
Eq.\ (\ref{Lambda phase transition}), we can see that it will be
negative and independent of the original cosmological constant
$\Lambda$ when the coupling constant $\lambda$ vanishes:
\begin{equation}
\Lambda _{eff} \longrightarrow - \frac{\mu ^2}{4\xi} \quad
\mbox{as} \quad \lambda \longrightarrow 0 ,
\end{equation}
which has been pointed out in our previous work \cite{Gu:2001rr}.
Consequently, for $\lambda =0$, the value of effective
cosmological constant $\Lambda_{eff}$, after the phase transition,
is controlled only by the parameters: $\mu ^2$ and $\xi$. This
particular feature might give some hint for dealing with the
cosmological constant problem. In addition, if we require a
comparatively small effective cosmological constant indicated by
various astronomical observations, for instance, the supernova
distance measurements, we need to fine tune the parameters: $\mu$
and $\lambda$, and the original cosmological constant $\Lambda$,
such that the condition
\begin{equation}
\frac{1}{\kappa} \Lambda \simeq \frac{\mu ^4}{4 \lambda} %
\label{fine tunning condition 2}
\end{equation}
is fulfilled. We note that the above condition, where the
non-minimal coupling with gravity is involved, is exactly the same
as the one in Eq.\ (\ref{fine tunning condition 1}), which is for
the case in a flat space-time. It is somewhat unexpected for us
that the fine-tuning condition is unchanged both qualitatively and
quantitatively even after introducing the non-minimal coupling.

\section*{Acknowledgements}
This work is supported in part by the National Science Council,
Taiwan, R.O.C. (NSC 89-2112-M-002 062) and by the CosPA project of
the Ministry of Education (MOE 89-N-FA01-1-4-3).

\end{document}